\documentclass[aps,superscriptaddress,eqsecnum,nofootinbib,showpacs,preprintnumbers,twocolumn]{revtex4-1}
\usepackage{longtable}
\usepackage{bm}
\usepackage{relsize}
\usepackage{amsfonts}
\usepackage{amsmath}
\usepackage{amssymb,epsf}
\usepackage{latexsym}
\usepackage{graphicx,epsfig}
\usepackage{amssymb}
\usepackage{float}
\usepackage{subfigure}
\usepackage{epstopdf}
\usepackage[colorlinks=true,citecolor=blue,linkcolor=blue,urlcolor=black]{hyperref}
\usepackage{dcolumn}
\usepackage{psfrag}
\usepackage{wrapfig}
\usepackage{makeidx}
\usepackage{epsf}
\usepackage{color}
\usepackage{multirow}
\usepackage{mathtools}

\def\be{\begin{equation}}
\def\ee{\end{equation}}
\def\bea{\begin{eqnarray}}
\def\eea{\end{eqnarray}}

\begin{document}

\title{$\Lambda$CDM-like evolution in Einstein-scalar-Gauss-Bonnet gravity}

\author{Miguel A. S. Pinto}
\email{mapinto@fc.ul.pt}
\affiliation{Instituto de Astrofísica e Ciências do Espaço, Faculdade de Ciências da Universidade de Lisboa, Edifício C8, Campo Grande, P-1749-016 Lisbon, Portugal}
\affiliation{Departamento de F\'{i}sica, Faculdade de Ci\^{e}ncias da Universidade de Lisboa, Edifício C8, Campo Grande, P-1749-016 Lisbon, Portugal}

\author{João Luís Rosa}
\email{joaoluis92@gmail.com}
\affiliation{Departamento de F\'isica Te\'orica, Universidad Complutense de Madrid, E-28040 Madrid, Spain}
\affiliation{Institute of Physics, University of Tartu, W. Ostwaldi 1, 50411 Tartu, Estonia}

\date{\today}

\begin{abstract}
In this work, we analyze the Einstein-scalar-Gauss-Bonnet (EsGB) theory of gravity in a cosmological context using the formalism of dynamical systems. We obtain the equations of motion of the theory and introduce an appropriate set of dynamical variables to allow for a direct comparison with the results from General Relativity (GR). We observe that the cosmological phase space features the same set of fixed points as in standard GR, i.e., radiation-dominated, matter-dominated, curvature-dominated, and exponentially-accelerated solutions independently of the values of the coupling function and the scalar field. Furthermore, the radiation-dominated fixed points are repellers and the exponentially accelerated fixed points are attractors in the phase space, thus allowing for cosmological solutions behaving qualitatively similar to the $\Lambda$CDM model, i.e., transitioning from a radiation-dominated phase into a matter-dominated phase, and later into a late-time cosmic acceleration phase supported by the scalar field potential. Following a reconstruction method through which we produce the cosmological solutions in the GR limit of the theory and introduce them into the general EsGB dynamical system, a numerical integration of the dynamical system shows that the EsGB theory provides cosmological solutions indistinguishable from those of the standard $\Lambda$CDM model, compatible with the current observations from the Planck satellite and weak-field solar system dynamics, while maintaining the scalar field and the coupling function finite and regular throughout the entire time evolution. 
\end{abstract}

\maketitle


\section{Introduction}\label{sec:intro}

The universe has been observed to be currently undergoing a phase of accelerated expansion, according to observations from type IA supernovae \cite{SupernovaCosmologyProject:1998vns,SupernovaSearchTeam:1998fmf} and the cosmic microwave background \cite{Planck:2018vyg}. In the context of General Relativity (GR) such a cosmological behavior requires the existence of an exotic fluid component known as dark energy \cite{Copeland:2006wr,Li:2011sd,Peebles:2002gy,Bamba:2012cp}. Although these dark energy models successfully explain the observations above, an alternative approach to address this issue without the necessity of recurring to unknown energy sources is the modification of the underlying gravity theory \cite{Clifton:2011jh,Capozziello:2011et,Nojiri:2017ncd,Nojiri:2010wj,Odintsov:2023weg}. The methods to extend GR are vast and multifaceted, from generalizations of the geometrical Lagrangian e.g. $f(R)$ gravity \cite{Sotiriou:2008rp,DeFelice:2010aj}, to the addition of extra fundamental fields \cite{DeFelice:2010jn} and geometrical invariants \cite{Fernandes:2022zrq}. Among these, a theory that has been scrutinized in different contexts is the Einstein-scalar-Gauss-Bonnet (EsGB) gravity.

The EsGB gravity arises in the compactified low-energy limit of string theory \cite{Gross:1986mw,Metsaev:1987zx,Zwiebach:1985uq,Cano:2021rey}. It consists of an extension of GR via the addition of a scalar field coupled to higher-order curvature terms through the Gauss-Bonnet invariant. This theory is particularly appealing for admitting hairy black-hole (BH) solutions \cite{Herdeiro:2015waa,Sotiriou:2013qea,Pani:2011gy,Maselli:2015tta}. These solutions can arise dynamically  through a process known as spontaneous scalarization \cite{Silva:2017uqg,Doneva:2017bvd,Cunha:2019dwb,Doneva:2022ewd}, which can have observational imprints in the gravitational wave signal if the scalarized compact object is part of a binary system \cite{Shiralilou:2020gah,Yagi:2011xp}. The properties of scalarized BHs in EsGB gravity have been studied including their stability \cite{Blazquez-Salcedo:2024rvb,Kleihaus:2023zzs,Antoniou:2022agj} and shadows \cite{Konoplya:2019fpy,Cunha:2016wzk}. In the context of cosmology, theories with spontaneous scalarization were shown to feature GR as a cosmological attractor \cite{Antoniou:2020nax}. The EsGB gravity has also been analyzed in different cosmological contexts \cite{Micolta-Riascos:2024vbm,Kanti:2015dra,Odintsov:2018zhw,Nojiri:2023mbo,Odintsov:2023lbb,Nojiri:2023mvi,MohseniSadjadi:2024ejb,Chakraborty:2018scm}, including through the formalism of dynamical systems  \cite{Dialektopoulos:2022kiv,Millano:2023gkt,Chatzarakis:2019fbn,Hussain:2024yee,
TerenteDiaz:2023iqk}.

The formalism of dynamical systems is one of the most versatile methods to analyze the cosmological phase space of a gravitational theory \cite{Bahamonde:2017ize}, leading to a wide plethora of applications in the framework of modified theories of gravity \cite{Odintsov:2017tbc,Carloni:2015jla,Alho:2016gzi,Carloni:2007eu,Rosa:2023qun,Carloni:2017ucm,Carloni:2009jc,Carloni:2015lsa,Carloni:2018yoz,Rosa:2019ejh,Carloni:2007br,Carloni:2013hna,Bonanno:2011yx,Goncalves:2023klv,Rosa:2024pzo,Rosa:2024pzo,Kaczmarek:2024quk}. However, the success of the method relies strongly on how the dynamical system and its quantities are constructed. Indeed, even though the dynamical system approach was already used in the context of EsGB gravity \cite{Dialektopoulos:2022kiv,Millano:2023gkt,Chatzarakis:2019fbn,Hussain:2024yee,
TerenteDiaz:2023iqk}, these analyses are heavily model-dependent, with explicit forms of the potential and coupling function being imposed in order to resolve the system. In this work, we aim to overcome these limitations of previous works by considering a more adequate definition of the dynamical system that allows for the analysis to be carried out without specifying a form of these functions \textit{a priori}, with the explicit forms being recovered posteriorly.

This manuscript is organized as follows. In Sec. \ref{sec:theory} we introduce the EsGB theory and obtain its equations of motion. Then, we introduce the background of a Friedmann-Lemaître-Robertson-Walker (FLRW) universe, and we obtain the equations of motion of the theory in this background. We also specify the matter distribution and the corresponding conservation equations. In Sec. \ref{sec:dynsys} we introduce a set of dynamical variables and the number of e-folds and we rewrite the equations of motion in the form of a dynamical system. We analyze the structure of the phase space including fixed points and phase diagrams, and we perform a numerical integration of the dynamical system to obtain cosmological solutions compatible with the current experimental measurements. Finally, we trace our conclusions in Sec. \ref{sec:concl}.

\section{Theoretical Framework}\label{sec:theory}

\subsection{Action and field equations}

The action functional $S$ that describes the EsGB gravity is given by
\begin{eqnarray}\label{eq:action}
    S &=& \frac{1}{2\kappa^2} \int_{\Omega}\sqrt{-g} \big[R - \frac{1}{2}g^{\mu\nu} \partial_\mu \phi\partial_\nu \phi -V(\phi) + \nonumber\\    
    &+&\alpha h(\phi) \mathcal{G}+2\kappa^2 \mathcal{L}_m(g_{\mu\nu},\chi) \big]d^4 x 
\end{eqnarray}
where $\kappa^2=8\pi G/c^4$, with $G$ the gravitational constant and $c$ the speed of light, $\Omega$ is the 4-dimensional spacetime manifold on which one defines a set of coordinates $x^\mu$, $g$ is the determinant of the metric $g_{\mu\nu}$, $R$ is the Ricci scalar, $\phi$ is a scalar field, $V\left(\phi\right)$ is the potential of the scalar field, $\alpha$ is a constant free parameter controlling the contribution of the Gauss-Bonnet term, $h\left(\phi\right)$ is a coupling function of $\phi$, $\mathcal G$ is the Gauss-Bonnet invariant defined as
\begin{equation}\label{eq:defGB}
    \mathcal G=R^2-4R_{\mu\nu}R^{\mu\nu}+R_{\mu\nu\sigma\rho}R^{\mu\nu\sigma\rho},
\end{equation}
where $R_{\mu\nu}$ is the Ricci tensor and $R_{\mu\nu\sigma\rho}$ is the Riemann tensor, $\mathcal L_m$ is the matter Lagrangian, and $\chi$ collectively denotes any matter fields. Equation \eqref{eq:action} depends explicitly on two independent quantities, namely the metric $g_{\mu\nu}$ and the scalar field $\phi$. The modified field equations for the EsGB gravity can be obtained by taking a variation of Eq. \eqref{eq:action} with respect to the metric $g_{\mu\nu}$ from which one obtains
\begin{eqnarray}\label{eq:field}
     && G_{\mu\nu}+ \frac{1}{2}g_{\mu\nu}\left[V\left(\phi\right)+\frac{1}{2}\nabla_{\sigma}\phi\nabla^{\sigma}\phi \right]-\frac{1}{2}\nabla_{\mu}\phi\nabla_{\nu}\phi
     \nonumber \\
    && + 4 \alpha\left(R_{\mu \sigma \nu \rho}+2 g_{\sigma[\nu} R_{\rho] \mu}+2 g_{\mu[\sigma} G_{\nu] \rho}\right) \nabla^\sigma \nabla^\rho h\left(\phi \right)
    \nonumber \\
    && = \kappa^2 T_{\mu \nu},
\end{eqnarray}
where $\nabla_\mu$ denotes the covariant derivatives, $G_{\mu\nu}\equiv R_{\mu\nu}-\frac{1}{2}g_{\mu\nu}R$ is the Einstein's tensor, $T_{\mu\nu}$ is the stress-energy tensor, defined in terms of the variation of the matter Lagrangian as
\begin{equation}\label{eq:deftab}
T_{\mu\nu}=-\frac{2}{\sqrt{-g}}\frac{\delta\left(\sqrt{-g}\mathcal L_m\right)}{\delta g^{\mu\nu}},
\end{equation}
and we have introduced the notation for index anti-symmetrization as $X_{[\mu\nu]}\equiv\frac{1}{2}\left(X_{\mu\nu}-X_{\nu\mu}\right)$. On the other hand, the equation of motion for the scalar field $\phi$ can be obtained by taking the variation of Eq. \eqref{eq:action} with respect to $\phi$, yielding
\begin{equation}\label{eq:eomphi}
    \Box\phi - V_{\phi} + \alpha h_{\phi}\mathcal{G}=0,
\end{equation}
where $\Box\equiv \nabla_\mu\nabla^\mu$ is the d'Alembert operator and we have introduced the notation $V_\phi\equiv dV/d\phi$ and $h_\phi\equiv dh/d\phi$. We note that, by taking the covariant derivative of Eq. \eqref{eq:field} and using Eq. \eqref{eq:eomphi}, one obtains the conservation equation
\begin{equation}\label{eq:conservation}
    \nabla_\mu T^{\mu\nu}=0,
\end{equation}
i.e., energy is conserved in EsGB gravity. 

\subsection{Geometry and matter distribution}

In this work, we aim to analyze the EsGB theory in a cosmological context. For this purpose, we assume that the spacetime is well described by a homogeneous and isotropic universe with some spacial curvature $k$. Thus, we adopt the FLRW line element in the usual spherical coordinates $x^\mu=\left(t,r,\theta,\varphi\right)$ in the form
\begin{equation}\label{eq:metric}
    ds^2 = -dt^2 +a^2\left(t\right)\left[\frac{dr^2}{1-kr^2}+r^2 d\Omega^2\right],
\end{equation}
where $a\left(t\right)$ is the scale factor of the universe, assumed to depend solely on the time coordinate $t$ as to preserve the homogeneity and isotropy of the spacetime, $k$ takes the values $k=\{-1, 0, 1\}$ for hyperbolic, flat, and spherical geometries respectively, and $d\Omega^2 =d\theta^2 + \sin^2{\theta} d\varphi^2$ is the line-element on the two-sphere. In what follows, it is useful to define the Hubble parameter $H$ as
\begin{equation}\label{eq:hubble}
    H = \frac{\dot{a}}{a},
\end{equation}
where a dot $(\dot{\ })$ denotes a derivative with respect to time. It is also useful in what follows to introduce the deceleration parameter, which is also a dimensionless function, defined as
\begin{equation}\label{eq:defQ}
    Q=-\frac{\Ddot{a}}{aH^2}.
\end{equation}

Regarding the matter components, we assume that the distribution of matter is well described by an isotropic relativistic perfect fluid with an energy density $\rho$ and a pressure $p$. Under these assumptions, the stress-energy tensor $T_{\mu\nu}$ takes the form
\begin{equation}\label{eq:em-fluid}
T_{\mu\nu}=\left(\rho+p\right)u_\mu u_\nu +p g_{\mu\nu},
\end{equation}
where $u_\mu$ is the 4-velocity vector of the fluid satisfying the normalization condition $u^\mu u_\mu=-1$. Furthermore, we assume that this relativistic fluid is constituted by two components: a pressureless dust component described by an equation of state $p_m=w_m\rho_m$, with $w_m=0$, and a radiation component described by an equation of state $p_r=w_r\rho_r$, with $w_r=\frac{1}{3}$. Furthermore, we focus our attention on the phases of the evolution of the universe for which the conditions are not conducive to the transformation of matter between these two components. This implies that the two components of the fluid are independently conserved, and thus Eq. \eqref{eq:conservation} takes the forms
\begin{eqnarray}\label{eq:consr}
    \dot{\rho_r}+4H\rho_r=0,\\ \label{eq:consm}
    \dot{\rho_m}+3H\rho_m=0.
\end{eqnarray}
Finally, we consider that the scalar field and the matter fields depend solely on the cosmological time, in order to preserve the homogeneity of space-time.
Under the assumptions outlined above, the field equations in Eq. \eqref{eq:field} feature two independent components, corresponding to the modified Friedmann and Raychaudhuri equations. These equations take the following forms
\begin{equation}\label{eq:cosmo1}
    3\left(H^2+\frac{k}{a^2}\right) \left(1+4 \alpha  H \dot{h}\right)=\kappa ^2\left(\rho_m+ \rho_{r}\right)
   +\frac{V}{2} + \frac{\dot{\phi}^2}{4},
\end{equation}
\begin{eqnarray}\label{eq:cosmo2}
&&-\frac{\kappa^2\rho_r}{3}+\frac{V}{2}-\frac{\dot\phi^2}{4}\\
&&=\left(H^2+\frac{k}{a^2}\right)\left(1+4\alpha\ddot{h}\right)-2QH^2\left(1+4\alpha H\dot{h}\right).\nonumber
\end{eqnarray}
On the other hand, the equation of motion for the scalar field $\phi$ given in Eq. \eqref{eq:eomphi} takes the form 
\begin{equation}\label{eq:cosmophi}
    \ddot{\phi} + 3H\dot\phi+ V_\phi + 24 \alpha  H^2 Q  \left(H^2+\frac{k}{a^2}\right)\frac{\dot{h}}{\dot{\phi}} = 0.
\end{equation}

The system of Eqs. \eqref{eq:consr} to \eqref{eq:cosmophi} consists of a system of five equations of which only four are linearly independent. This can be shown by taking a derivative of Eq. \eqref{eq:cosmo1} and then using Eqs. \eqref{eq:consr}, \eqref{eq:consm}, \eqref{eq:cosmo1}, \eqref{eq:cosmo2} and \eqref{eq:cosmophi} to eliminate the terms $\dot\rho_r$, $\dot{\rho_m}$, $\dot a$, $\ddot a$, and $\ddot \phi$, from which one obtains an identity. Thus, we have a system of four linearly independent equations to be solved for six unknown functions, namely $a$, $\phi$, $\rho_r$, $\rho_m$, $V$, and $h$. This is an underdetermined system and additional constraints can be imposed to achieve determination.

\section{Dynamical system approach}\label{sec:dynsys}

\subsection{Dynamical variables and equations}

In order to perform an analysis using the dynamical system approach, it is necessary to introduce a set of dimensionless dynamical variables to describe the quantities of interest in the system, as well as a dimensionless time coordinate. In this case, we define the following set of dynamical variables \footnote{Note that the set of dynamical variables defined in Eq. \eqref{eq:dynvars} is potentially problematic for cosmological solutions in which $H=0$, i.e., for static universes, or for recollapsing universes. Although these cases are mathematically acceptable, they constitute cosmological solutions that are not well-motivated from a physically realistic perspective, and thus we do not consider an alternative set of dynamical variables that could address those solutions. Nevertheless, this could be achieved through the definition of an additional dynamical variable $\mathcal H = H / H_0$, for some constant $H_0$, and by replacing $H$ by $H_0$ in Eqs. \eqref{eq:dynvars}.}
\begin{equation}
\begin{gathered}\label{eq:dynvars}
K=\frac{k}{a^2 H^2}, \quad \Omega_r=\frac{8 \pi \rho_r}{3 H^2}, \quad \Omega_m=\frac{8 \pi \rho_m}{3 H^2}, \\
\Phi=\phi,\quad \Theta=\frac{\dot{\phi}}{H},\quad \Psi= H\dot{h}, \quad U=\frac{V}{6 H^2}.
\end{gathered}
\end{equation}
In addition, we also introduce the following auxiliary variable
\begin{equation}
\label{Aux_variable}
    \Gamma = \frac{\ddot{\phi}}{H^2},
\end{equation}
which is useful in the follow-up analysis. Furthermore, as a dimensionless time coordinate, we take the number of e-folds $N$, described by
\begin{equation}\label{eq:defN}
    N \equiv \log\left(\frac{a}{a_0}\right),
\end{equation}
where $a_0$ is the present value of the scale factor, such that $N=0$ represents the present time. The derivatives with respect to $t$ present in the equations of motion can thus be converted into derivatives with respect to $N$ through the chain rule
\begin{equation}\label{eq:chainN}
X^{\prime} \equiv \frac{d X}{d N}=\frac{1}{H} \frac{d X}{d t}=\frac{\dot{X}}{H},
\end{equation}
where a prime $(')$ denotes a derivative with respect to $N$. Introducing the definitions of Eqs. \eqref{eq:dynvars} and \eqref{eq:defN} into the equations of motion in Eqs. \eqref{eq:consr} to \eqref{eq:cosmo2}, one obtains a constraint equation plus a set of dynamical equations for the dynamical variables $\Omega_r$, $\Omega_m$, and $\Psi$. A dynamical equation for the variable $U$ can be obtained through the introduction of the definitions above (including Eq. \eqref{Aux_variable}) into the chain rule $\dot V=V_\phi \dot\phi$ and using Eq. \eqref{eq:cosmophi} to eliminate the term $V_\phi$. 
Finally, the dynamical equations for the variables $K$, $\Phi$ and $\Theta$ can be obtained by taking directly a derivative of these variables with respect to $N$. The set of dynamical equations obtained through this method takes the form
\begin{equation}\label{eq:con1}
    (1+K) \left(1+4 \alpha  \Psi\right)=\Omega_m + \Omega
_r+U+\frac{\Theta^2}{12}.
\end{equation}
\begin{eqnarray}\label{eq:con2}
&&\left(1+K\right)\left\{1+4\alpha\left[\left(1-\frac{1-K}{1+K}Q\right)\Psi+\Psi'\right]\right\}=\nonumber \\
&&=2Q+3U-\Omega_r - \frac{\Theta^2}{4}
\end{eqnarray}
\begin{equation}\label{eq:dynK}
     K' = 2 K  Q,
\end{equation}
\begin{equation}\label{eq:dynm}
     \Omega_m' = \Omega_m \left(2Q-1 \right),
\end{equation}
\begin{equation}\label{eq:dynr}
    \Omega_r' =  2 \Omega_r\left(Q-1\right),
\end{equation}
\begin{equation}\label{eq:dynU}
U'=2U\left(Q+1\right)-\frac{\Theta^2}{2}-\frac{1}{6}\Theta\Gamma - 4\alpha Q \left(1+K\right) \Psi,
\end{equation}
\begin{equation}
\label{eq:dynPhi}
    \Phi'=\Theta,
\end{equation}
\begin{equation}
\label{eq:dynTheta}
\Theta'=\left(1+Q\right)\Theta + \Gamma.
\end{equation}
These equations fully describe the cosmology of EsGB gravity in the form of a dynamical system. We note that the self consistency previously identified for the equations of motion before the introduction of the dynamical variables still holds at this point, i.e., of the eight equations in Eqs. \eqref{eq:con1} to \eqref{eq:dynTheta} only seven are linearly independent, and thus a complete solution of the dynamical system can be obtained even if e.g. Eq. \eqref{eq:con2} is discarded. 

\subsection{Phase space}

The dynamical system described by Eqs. \eqref{eq:con1} to \eqref{eq:dynTheta} presents a total of three invariant submanifolds, corresponding to the submanifolds $K=0$, $\Omega_m=0$, and $\Omega_r=0$. This implies that any global property of the phase space, e.g., a global attractor, must lie in the intersection of these three invariant submanifolds. Furthermore, the system features five sets of fixed points that are summarized in Table \ref{tab:fixed}. The points $\mathcal{A}$ correspond to radiation dominated solutions, points $\mathcal{B}$ are matter dominated solutions, points $\mathcal{C}$ are exponentially accelerated solutions, and both points $\mathcal{D}$ and $\mathcal{E}$ are curvature dominated solutions, with $\mathcal{D}$ having a direct GR counterpart. Note that in the limit $\alpha\to 0$, the sets of points $\mathcal{A}-\mathcal{D}$ reduce to the set of fixed points in GR, while $\mathcal{E}$ becomes singular.

In order to clarify the stability of the fixed points, we resort to linear stability theory. For more details on this procedure, we recommend the Appendix of Ref. \cite{Goncalves:2023klv}. We consider the full dynamical system, Eqs.\eqref{eq:con1} to \eqref{eq:dynTheta}, and we obtain the corresponding Jacobian matrix and its eigenvalues, for all fixed points previously found. A summary of the eigenvalues for each fixed point is given in Table \ref{tab:Stability}. These results indicate that points $\mathcal A$ are unstable, thus representing a repeller in the phase space, and points $\mathcal B$, $\mathcal{D}$, and $\mathcal{E}$ are saddle points. The fixed points $\mathcal C$ present two possible distinct behaviors: if the condition $\alpha \Psi < -1/2$ is satisfied, these fixed points are stable, thus consisting of attractors in the phase space. Furthermore, since these points lie at the intersection of the three invariant submanifolds $K=0$, $\Omega_r=0$, and $\Omega_m=0$, these attractors are global attractors in the phase space. Note that in the limit $\alpha\to0$, even though it does not satisfy the inequality above, the fixed points $\mathcal C$ still correspond to attractors in the phase space. This happens because in this limit, the dynamical variable $\Psi$, along which the points $\mathcal C$ change their stability behavior, is removed from the dynamical system.

In addition, for the sake of completeness and to make it easier to visualize the phase space, we perform projections into the invariant submanifolds of the dynamical system in the GR limit, i.e., considering $\Phi=\Phi_0 = $ constant, $\Theta = 0$, $\Psi=0$. We define the projections $M_i$ as projections of the dynamical system in which the variable $K$ has been removed from the system using the constraint equation in Eq. \eqref{eq:con1}, with $M_1$ corresponding to a projection into $\Omega_r=0$, $M_2$ corresponding to a projection into $\Omega_m=0$, and $M_3$ corresponding to a projection into $U=0$\footnote{We emphasize that the submanifold $U=0$ is only invariant for if $\Phi=\Phi_0 =$ constant. Nevertheless, given that the fixed points do not depend explicitly on the dynamical variable $\Phi$, the assumption $\Phi=\Phi_0$ does not qualitatively alter the structure of the phase space.}. We also define the projections $N_i$ as projections of the dynamical system in which the variable $U$ has been removed from the system using the constraint equation in Eq. \eqref{eq:con1}, with $N_1$ corresponding to a projection into $\Omega_r=0$, $N_2$ corresponding to a projection into $\Omega_m=0$, and $N_3$ corresponding to a projection into $K=0$.

\begin{figure*}
    \centering
    \includegraphics[scale=0.6]{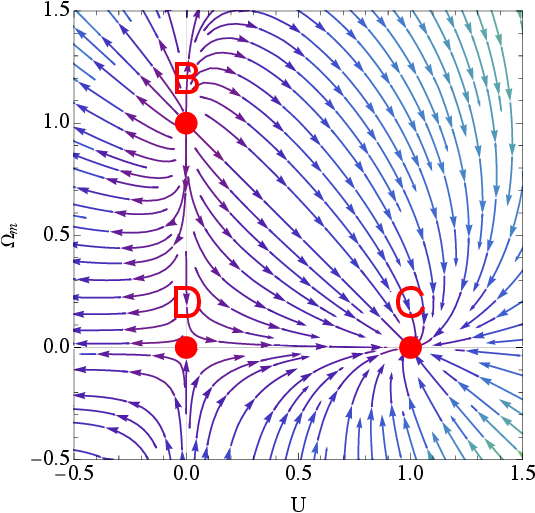}\qquad
    \includegraphics[scale=0.6]{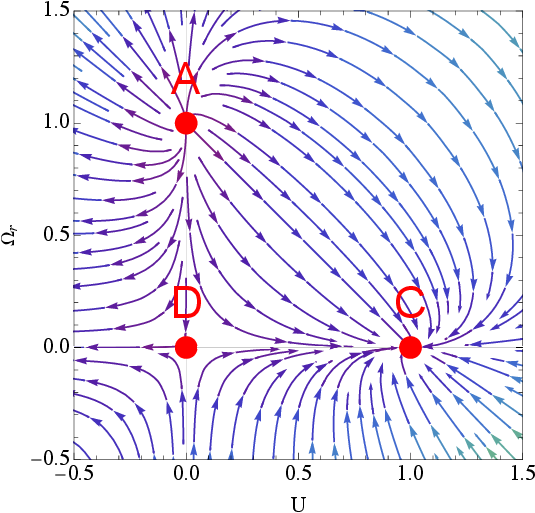}\qquad
    \includegraphics[scale=0.6]{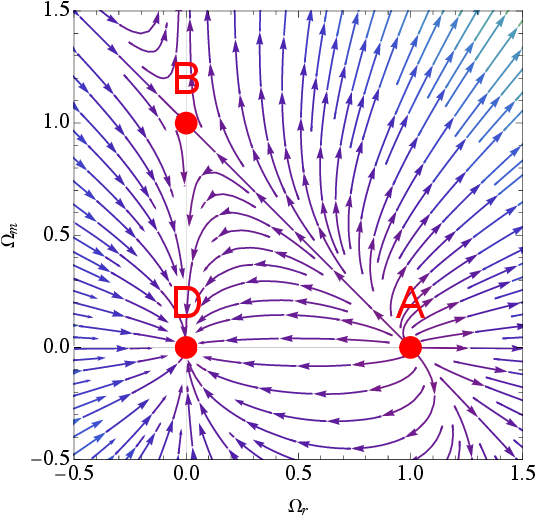}
    \caption{Streamplots of the cosmological phase space for the dynamical system given in Eqs. \eqref{eq:con1} to \eqref{eq:dynTheta}  projected into the invariant submanifolds $M_1$ (left panel), $M_2$ (middle panel), and $M_3$ (right panel). The stability analysis of the fixed points represented is given in Table \ref{tab:StabilityGR1}.}
    \label{fig:StreamPlots1}
\end{figure*}

\begin{figure*}
    \centering
    \includegraphics[scale=0.6]{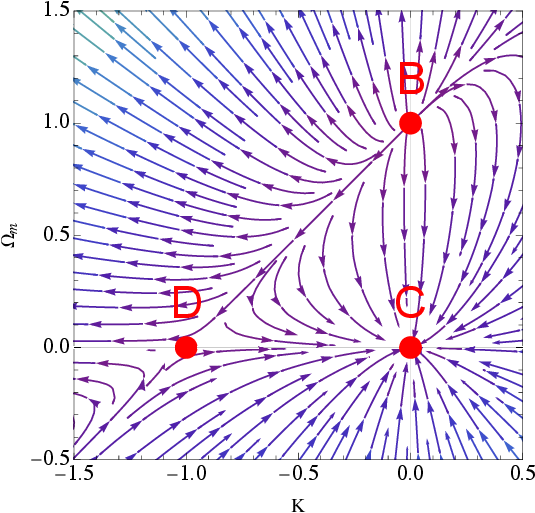}\qquad
    \includegraphics[scale=0.6]{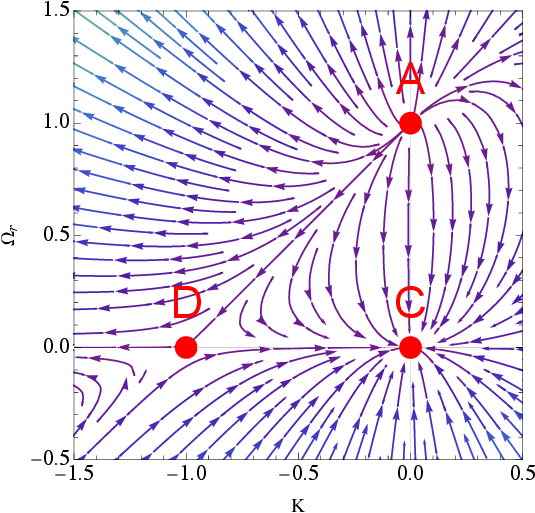}\qquad
    \includegraphics[scale=0.6]{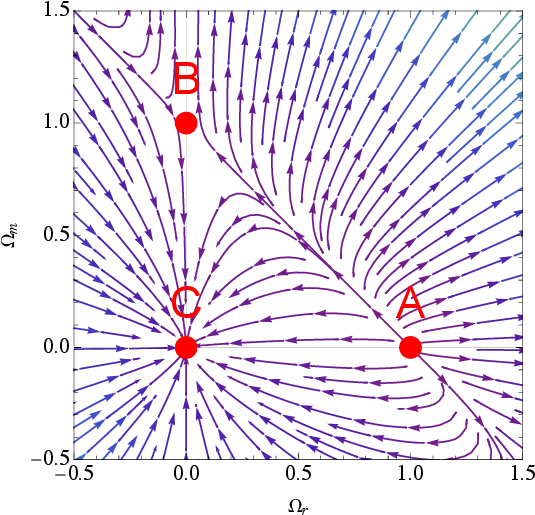}
    \caption{Streamplots of the cosmological phase space for the dynamical system given in Eqs. \eqref{eq:con1} to \eqref{eq:dynTheta}  projected into the invariant submanifolds $N_1$ (left panel), $N_2$ (middle panel), and $N_3$ (right panel). The stability analysis of the fixed points represented is given in Table \ref{tab:StabilityGR2}.}
    \label{fig:StreamPlots2}
\end{figure*}

A summary of the fixed points visible from each of the projections $M_i$ and $N_i$, alongside their eigenvalues and resultant stability, is given in Tables \ref{tab:StabilityGR1} and \ref{tab:StabilityGR2}, respectively, whereas the corresponding phase space trajectories are shown in Figs. \ref{fig:StreamPlots1} and \ref{fig:StreamPlots2}, respectively. These results show that, independently of the projection taken, point $\mathcal A$ always behaves as a repeller, whereas point $\mathcal C$ always behaves as an attractor, thus confirming the unstable and stable characters of these points, respectively. On the other hand, the fixed point $\mathcal B$ behaves as a repeller as seen from the projections $M_1$ and $N_1$, or a saddle as seen from the projections $M_3$ and $N_3$, whereas point $\mathcal D$ behaves either as a saddle as seen from the projections $M_1$, $M_2$, $N_1$, and $N_2$, or as an attractor, as seen from the projection $M_3$. These results are in accordance with the ones found for the total system (see Table \ref{tab:Stability}), i.e., that points $\mathcal B$ and $\mathcal D$ are saddle points as seen from the whole phase space dynamical system as a whole. Finally, we note that, in the projection $N_3$, one observes trajectories in the phase space that emerge from point $\mathcal A$, approach point $\mathcal B$, and finally evolve towards point $\mathcal C$, which hints at the possibility of having a single cosmological solution evolving from a radiation-dominated phase into a matter-dominated phase and then into a late-time cosmologically accelerated phase. 

\begin{table*}
\begin{tabular}{c|c c c c c c c c} 
 & $K$ & $\Omega_r$ & $\Omega_m$ & $U$ & $\Theta$ & $\Psi$ & $\Gamma$ & $Q$ \\
\hline $\mathcal{A}$ & 0 & $1 +3\alpha \Psi$ & 0 & $\alpha \Psi$ & 0 & ind. & 0 & 1 \\
$\mathcal{B}$ & 0 & 0 & $\frac{1}{3}\left(3+10 \alpha \Psi\right)$ & $\frac{2}{3}\alpha \Psi$ & 0 & ind. & 0 & $\frac{1}{2}$ \\
$\mathcal{C}$ & 0 & 0 & 0 & $1+4\alpha \Psi$ & 0 & ind. & 0 & $\frac{-1-4 \alpha \Psi}{1 + 2\alpha \Psi}$ \\
$\mathcal{D}$ & -1 & 0 & 0 & 0 & 0 & ind. & 0 & 0 \\
$\mathcal{E}$ & ind. & 0 & 0 & 0 & 0 & $- \frac{1}{4\alpha}$ & 0 & 0
\end{tabular}
\caption{Fixed points arising from the dynamical system of Eqs. \eqref{eq:con1} to \eqref{eq:dynTheta}. The tag ”ind.”
indicates an independent quantity.}
\label{tab:fixed}
\end{table*}

\begin{table*}
    \centering
    \begin{tabular}{c| c | c | c| c| c| c| c | c}
         & $\lambda_1$ & $\lambda_2$ & $\lambda_3$ & $\lambda_4$ & $\lambda_5$ & $\lambda_6 $ & $\lambda_7$ & Stability\\ 
        \hline $\mathcal{A}$ & 0 & 0 & 2 & 2 & 1 & 1 & 3 & Repeller\\ 
        $\mathcal{B}$ & 0 & -1 & 1 & 3/2 & 0 & 0 & 5/2 & Saddle \\ 
        $\mathcal{C}$ &  0 & $  \frac{2}{1+2\alpha \Psi}$ - 6 & $\frac{2}{1+2\alpha \Psi} -4$ & $\frac{1}{1+2\alpha \Psi} - 1$ & $\frac{2}{1+2\alpha \Psi} - 5$ & -$\frac{5}{2}+\frac{\frac{3}{2}-2-5\alpha\Psi}{1+2\alpha\Psi}$& -$\frac{5}{2}+\frac{\frac{3}{2}+2+5\alpha\Psi}{1+2\alpha\Psi}$ & Attractor or Saddle \\ 
        $\mathcal{D}$ & 0 & -2 & 0 & 1 & -1 & -1 & 2 & Saddle \\ 
        $\mathcal{E}$ & 0 & -2 & 0 & 1 & -1 & -1 & 2 & Saddle \\ 
    \end{tabular}
    \caption{Eigenvalues $\lambda_i$ and stability of the fixed points of the dynamical system in Eqs. \eqref{eq:con1} to \eqref{eq:dynTheta}.}
    \label{tab:Stability}
\end{table*}

\begin{table}
    \centering
    \begin{tabular}{c|c|c|c|c}
         & $\mathcal A$ & $\mathcal B$ & $\mathcal C$ & $\mathcal D$ \\ \hline
        $M_1$ & X & $\begin{matrix}\lambda_1=3\\ \lambda_2=1\end{matrix}$ (R) & $\begin{matrix}\lambda_1=-3\\ \lambda_2=-2\end{matrix}$ (A) & $\begin{matrix}\lambda_1=2\\ \lambda_2=-1\end{matrix}$ (S)  \\ \hline 
        $M_2$ & $\begin{matrix}\lambda_1=4\\ \lambda_2=2\end{matrix}$ (R) & X & $\begin{matrix}\lambda_1=-4\\ \lambda_2=-2\end{matrix}$ (A) & $\begin{matrix}\lambda_1=-2\\ \lambda_2=2\end{matrix}$ (S)  \\ \hline 
        $M_3$ & $\begin{matrix}\lambda_1=2\\ \lambda_2=1\end{matrix}$ (R) & $\begin{matrix}\lambda_1=-1\\ \lambda_2=1\end{matrix}$ (S) & X & $\begin{matrix}\lambda_1=-2\\ \lambda_2=-1\end{matrix}$ (A)  \\ \hline 
    \end{tabular}
    \caption{Eigenvalues $\lambda_i$ and stability of the fixed points of the dynamical system in Eqs. \eqref{eq:con1} to \eqref{eq:dynTheta} projected into the invariant submanifolds $\Omega_r=0$ ($M_1$), $\Omega_m=0$ ($M_2$), and $U=0$ ($M_3$) in the GR limit.  (A) stands for attractor, (R) stands for repeller, and (S) stands for saddle. An X indicates that the fixed point is not visible in the given projection.}
    \label{tab:StabilityGR1}
\end{table}

\begin{table}
    \centering
    \begin{tabular}{c|c|c|c|c}
         & $\mathcal A$ & $\mathcal B$ & $\mathcal C$ & $\mathcal D$ \\ \hline
        $N_1$ & X & $\begin{matrix}\lambda_1=3\\ \lambda_2=1\end{matrix}$ (R) & $\begin{matrix}\lambda_1=-3\\ \lambda_2=-2\end{matrix}$ (A) & $\begin{matrix}\lambda_1=2\\ \lambda_2=-1\end{matrix}$ (S)  \\ \hline 
        $N_2$ & $\begin{matrix}\lambda_1=4\\ \lambda_2=2\end{matrix}$ (R) & X & $\begin{matrix}\lambda_1=-4\\ \lambda_2=-2\end{matrix}$ (A) & $\begin{matrix}\lambda_1=-2\\ \lambda_2=2\end{matrix}$ (S)  \\ \hline 
        $N_3$ & $\begin{matrix}\lambda_1=4\\ \lambda_2=1\end{matrix}$ (R) & $\begin{matrix}\lambda_1=3\\ \lambda_2=-1\end{matrix}$ (S) & $\begin{matrix}\lambda_1=-4\\ \lambda_2=-3\end{matrix}$ (A) & X  \\ \hline 
    \end{tabular}
    \caption{Eigenvalues $\lambda_i$ and stability character of the fixed points of the dynamical system in Eqs. \eqref{eq:con1} to \eqref{eq:dynTheta} projected into the invariant submanifolds $\Omega_r=0$ ($N_1$), $\Omega_m=0$ ($N_2$), and $K=0$ ($N_3$) in the GR limit. (A) stands for attractor, (R) stands for repeller, and (S) stands for saddle. An X indicates that the fixed point is not visible in the given projection.}
    \label{tab:StabilityGR2}
\end{table}

\subsection{Full numerical integration}

To verify whether cosmological solutions qualitatively similar to the $\Lambda$CDM model in GR and consistent with the current cosmological observations exist in EsGB gravity, in this section, we implement a reconstruction method and perform a numerical integration of the dynamical system subjected to appropriate initial conditions.

\begin{figure*}
	\centering
\includegraphics[scale=0.95]{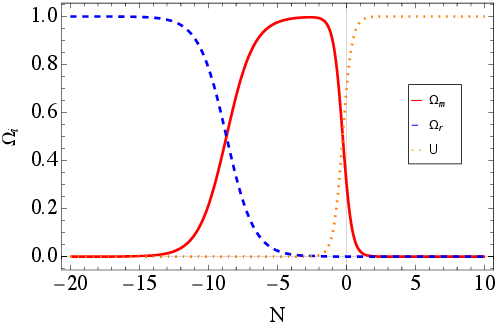}
\includegraphics[scale=0.90]{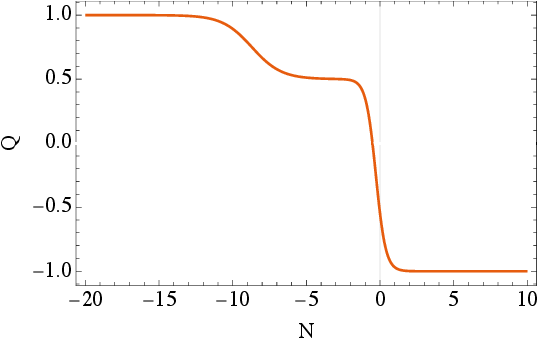}
	\caption{Density parameters $\Omega_r$, $\Omega_m$, and $U$ (left panel) and deceleration parameter $Q$ (right panel) as a function of the number of e-folds $N$, obtained through a numerical integration of the dynamical system defined by Eqs. \eqref{eq:con1} to \eqref{eq:dynTheta} in the GR limit $\Phi=\Phi_0=$ constant, $\Theta=0$, and $\Psi = 0$.}
 \label{fig:parameters}
\end{figure*}

We start by imposing the GR limit, i.e., $\Phi=\Phi_0 = $ constant, $\Theta = 0$, $\Psi=0$ for which the dynamical system in Eqs. \eqref{eq:dynK} to \eqref{eq:dynTheta} reduces to the corresponding dynamical system in GR, and Eqs.\eqref{eq:con1} and \eqref{eq:con2} become constraint equations, i.e., without derivatives of the dynamical variables. These two constraint equations allow us to remove two dynamical quantities from the system, e.g. one can use Eq. \eqref{eq:con2} to remove $Q$ from the system, and then use Eq. \eqref{eq:con1} to remove $U$ from the system. Once the system has been numerically resolved, one can always recover the solutions for $Q$ and $U$ from the same equations. Furthermore, given that the universe is observed to be approximately flat \cite{Planck:2018vyg}, we project the dynamical system into the invariant submanifold $K=0$. The remaining three dynamical equations for $\Omega_m$ and $\Omega_r$ can then be numerically integrated under the initial conditions $\Omega_r\left(0\right)=5\times 10^{-5}$ and $\Omega_m\left(0\right)=0.3$, consistent with the current cosmological observations. Introducing these solutions back into Eq. \eqref{eq:con1} one obtains the solution for $U$ and verifies that $U\left(0\right)=0.69995$, i.e., the potential $U$ effectively plays the role of dark energy at present times, and finally inserting these solutions into Eq. \eqref{eq:con2} one obtains the solutions for $Q$ and verifies that $Q\left(0\right)=-0.5499$, also consistent with the current cosmological observations. The solutions for $\Omega_r$, $\Omega_m$, $U$, and $Q$ obtained through this procedure are shown in Fig. \ref{fig:parameters}. As expected given that these solutions were obtained in the GR limit, we observe that the universe evolves from an early-time radiation-dominated phase with $\Omega_r\simeq 1$ and $Q\simeq 1$, transitions into a matter-dominated phase with $\Omega_m\simeq 1$ and $Q\simeq 0.5$, and finally transitions into a late-time cosmologically accelerated phase with $U\simeq 1$ and $Q\simeq -1$.

The cosmological solutions obtained previously correspond to the $\Lambda$CDM model in GR. Now, we are interested in verifying if these solutions are compatible with EsGB gravity and, if so, what is the behavior of the scalar field $\Phi$, its derivative, $\Theta$, and the function $\Psi$ that allows for that compatibility. 

\begin{figure*}[htbp]
	\centering
\includegraphics[scale=0.70]{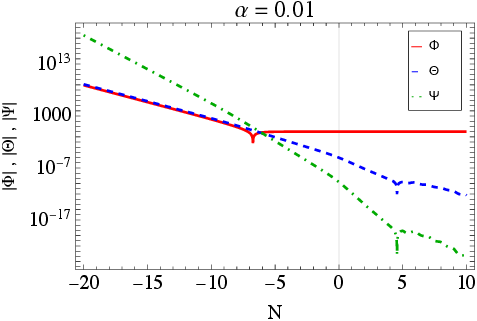}\quad
\includegraphics[scale=0.70]{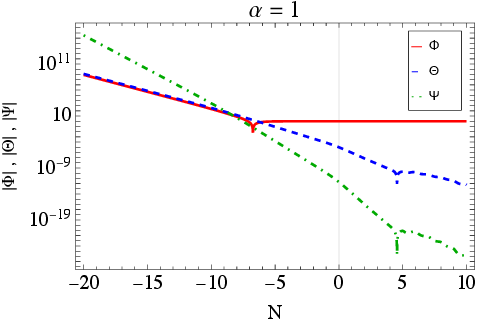}\quad
\includegraphics[scale=0.70]{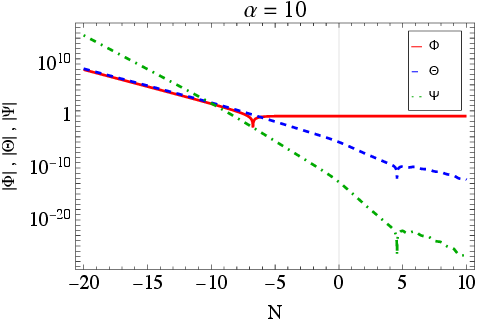} \\
\includegraphics[scale=0.70]{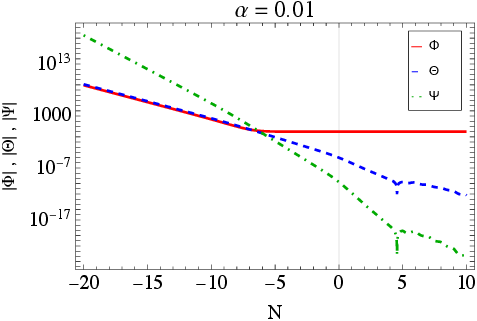}\quad
\includegraphics[scale=0.70]{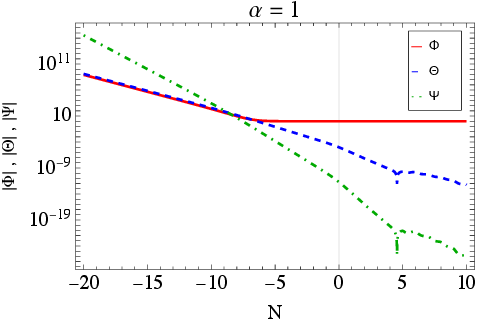}\quad
\includegraphics[scale=0.70]{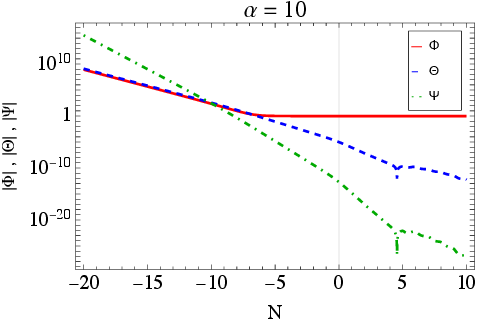}
	\caption{Solutions for $\Phi$, $\Theta$ and $\Psi$ as functions of the number of e-folds $N$ obtained from solving the dynamical system of Eqs. \eqref{eq:con1} to \eqref{eq:dynTheta} with the imposition of the solutions for $\Omega_r$, $\Omega_m$, $U$, and $Q$ from Fig. \ref{fig:parameters}. The initial conditions considered are $\Theta\left(0\right)=10^{-5}$ (top row), $\Theta\left(0\right)=-10^{-5}$ (bottom row) and $\Phi\left(0\right)=1$, and the values of the coupling constant adopted are $\alpha=0.01$ (left column), $\alpha=1$ (middle column), and $\alpha=10$ (right column).}
 \label{fig:solutions}
\end{figure*}

Invoking the previously mentioned linear dependence of the dynamical system, we discard Eq. \eqref{eq:con2} from the analysis without loss of information, and introduce the solutions obtained for $\Omega_m$, $\Omega_r$, $U$, and $Q$ into the dynamical system of Eq. \eqref{eq:con1} and Eqs. \eqref{eq:dynK} to \eqref{eq:dynTheta}. By doing so, Eq. \eqref{eq:con1} becomes an algebraic equation for the variables $\Theta$ and $\Psi$. Then, we use Eq. \eqref{eq:dynTheta} to eliminate the auxiliary variable $\Gamma$ from the system, thus turning Eq. \eqref{eq:dynU} into a first-order differential equation for $\Theta$. The solution for $\Theta$ is then introduced into Eqs. \eqref{eq:con1} and \eqref{eq:dynPhi} in order to obtain a solution for $\Psi$ and $\Phi$, respectively. To solve these equations numerically, it is necessary to impose a value on the coupling constant $\alpha$ and to provide initial conditions for $\Phi(0)$ and $\Theta(0)$. Note that an initial condition for $\Psi(0)$ is not required, given that $\Psi'$ only appears in Eq. \eqref{eq:con2}, which has been removed from the system. Given that, at present times, the weak field solar system dynamics are well modeled by the GR limit, any constant value for $\Phi(0)$ with $|\Theta(0)|\ll 1$ is compatible with this observation and, thus, represents a physically well-motivated choice for the present conditions. The solutions obtained for different combinations of $\alpha$, $\Phi(0)$, and $\Theta(0)$ are plotted in Fig. \ref{fig:solutions}.

The solutions for $\Phi$ and $\Psi$ present a few noteworthy features. First, one observes that $\Phi$ decays exponentially during the radiation dominated phase and transitions into a constant behavior in the transition to the matter dominated phase, a behavior that holds all the way to present and future times. On the other hand, $\Psi$ decays exponentially both during the radiation and matter dominated phases, although with different decaying rates as can be observed by a change in the slope of the curve, and eventually starts oscillating around zero when the transition to a late-time cosmic acceleration occurs. Note also that if $\alpha>0$, this implies that $\Psi<0$ at early times, whereas if $\alpha<0$, we have $\Psi>0$ at early times. Nevertheless, for both cases, $|\Psi|$ remains the same. Furthermore, if $\Phi(0)$ and $\Theta(0)$ have opposite signs, $\Phi$ remains always positive if $\Phi(0)>0$ and always negative if $\Phi(0)<0$, whereas if $\Phi(0)$ and $\Theta(0)$ have the same sign, $\Phi(0)>0$ implies that $\Phi<0$ a early times and changes sign during the transition from radiation to matter domination, and vice versa for $\Phi(0)<0$. Furthermore, the sign of $\alpha$ does not affect the behavior of $\Phi$ and $\Theta$. Nevertheless, independently of the choice of initial conditions and parameter $\alpha$, all quantities $\Phi$, $\Theta$, and $\Psi$ remain finite and regular throughout the entire cosmological evolution, thus indicating that cosmological solutions behaving in a qualitatively similar manner as the $\Lambda$CDM model in GR are attainable in EsGB gravity, with the scalar field $\phi$ being of particular importance at early times and the potential $V\left(\phi\right)$ playing the role of a cosmological constant at late times.

\subsection{Solution for $V\left(\phi\right)$ and $h\left(\phi\right)$}
Now that we have confirmed EsGB gravity accommodates a $\Lambda$CDM-like cosmological evolution while keeping $\Phi$, $\Theta$ and $\Psi$ regular the entire evolution, it is desirable to find what kind of potential $V\left(\phi\right)$ and coupling function $h\left(\phi\right)$ are compatible with such a solution. For this task, we write the deceleration parameter $Q$ in terms of the Hubble parameter and its derivative with respect to the cosmological time
\begin{equation}
    Q=-\frac{\dot{H}}{H^2} -1.
\end{equation}
Additionally, we introduce an auxiliary Hubble function in terms of the number of e-folds, $\mathcal{H}\left(N\right)=H\left(t\right)$ such that, in the space of the number of e-folds, the previous equality renders
\begin{equation}
\label{eq:Hnovo}
    Q=-\frac{\mathcal{H}'}{\mathcal{H}}-1.
\end{equation}
In order to obtain a numerical solution for $V\left(\phi\right)$, it is necessary first to obtain solutions for $U\left(\phi\right)$ and $H\left(\phi\right)$, see Eq. \eqref{eq:dynvars}. Moreover, if one takes the definition of $\Psi$ and applies the chain rule $\dot{h}=h_\phi \dot{\phi}$, where $h_{\phi}$ is the derivative of $h$ with respect to $\phi$, one can $\Psi$ as $\Psi=H^2h_{\phi}\Theta$. Hence, to obtain an explicit solution for $h\left(\phi\right)$, it is necessary to find solutions for $\Psi(\phi)$, $\Theta(\phi)$, and $H(\phi)$.

\begin{figure*}[htbp]
	\centering
\includegraphics[scale=0.95]{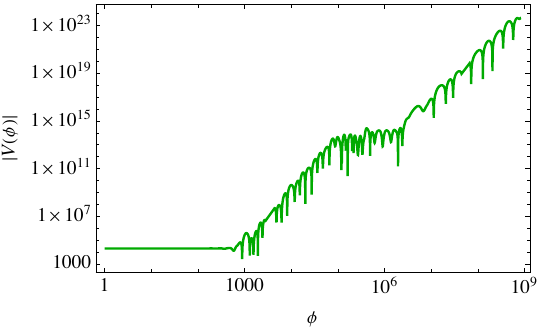}\quad
\includegraphics[scale=0.95]{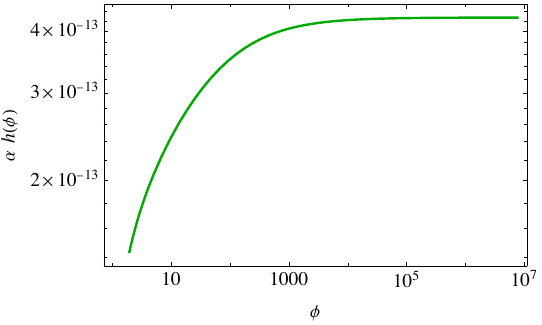}\quad
	\caption{Numerical solutions for $V\left(\phi\right)$ (left panel) and for $\alpha h\left(\phi\right)$ (right panel). The initial conditions considered are $\Theta\left(0\right)=-10^{-5}$ and $\Phi\left(0\right)=1$.}
 \label{fig:solutionsVh}
\end{figure*}

In the prior section, we found numerical solutions for $U(N)$, $\Phi(N)$, $\Theta(N)$, and $\Psi(N)$. Additionally, by adopting the initial condition $\mathcal{H}(0)=67.4$ compatible with the Planck data \cite{Planck:2018vyg}, one obtains a numerical solution for $\mathcal{H}(N)$ through Eq. \eqref{eq:Hnovo}. Then, since $\Phi(N)$ is a continuous and injective function, it can always be inverted into $N\left(\Phi\right)$, which is equivalent to $N\left(\phi\right)$ according to the definition of $\Phi$. Therefore, by using $N(\phi)$, it is possible to transform the previously found $U(N)$, $\Theta(N)$ and $\Psi(N)$ into the desired $U\left(\phi\right)$, $\Theta\left(\phi\right)$ and $\Psi\left(\phi\right)$, respectively. The same reasoning can be applied to $\mathcal{H}(N)$, where we can obtain $H(\phi)$ by taking into account $\mathcal{H}=H$. By adopting this procedure, one can obtain numerical solutions for both the scalar field potential $V(\phi)$, and for $ h(\phi)$. The solutions for $V(\phi)$ and $\alpha h(\phi)$ are plotted in Fig. \ref{fig:solutionsVh}, for the initial conditions $\Theta\left(0\right)=-10^{-5}$ and $\Phi\left(0\right)=1$.

Our results show that the resulting potential $V(\phi)$ is independent of the coupling constant $\alpha$, which means that its behavior is independent of the coupling between the scalar field and the Gauss-Bonnet invariant. Secondly, although the solutions $U(N)$ and $Q(N)$ correspond to the $\Lambda$CDM ones, $V(\phi)$ is not an overall cosmological constant but rather a sort of dynamical dark energy component. This occurs because $N(\phi)$ is not a constant value, thus rendering a non-trivial dependence on $\phi$. Nonetheless, in the limit $\phi\to1$, i.e., at present and future times ($N\to0$), the potential becomes effectively a cosmological constant. Finally, one observes that the potential becomes symmetrical with respect to the vertical axis $\phi=\Phi_0$ when the signs of either $\Phi(0)$ or $\Theta(0)$ are switched.

Concerning $h\left(\phi\right)$, and unlike it happens for $V\left(\phi\right)$, it depends explicitly on the value of $\alpha$. For symmetrical values of $\alpha$, $h(\phi)$ is symmetrical with respect to the horizontal axis. Furthermore, and similarly to what happens with the scalar field potential, switching the sign of either $\Phi(0)$ or $\Theta(0)$ causes $h(\phi)$ to be symmetrical with respect to the vertical axis $\phi=\Phi_0$. Nonetheless, $\alpha h\left(\phi\right)$ remains unchanged, which means that the coupling with the Gauss-Bonnet invariant is unique.

\section{Summary and Discussion}
\label{sec:concl}

In this work, we have analyzed the EsGB theory of gravity in a cosmological context through the use of the dynamical system approach. We have obtained the equations of motion of the theory and we have written these equations in a dimensionless form via the introduction of an appropriate set of dimensionless dynamical variables and dimensionless time coordinate (the number of e-folds $N$). Unlike previous works, the appropriateness of the set of dynamical variables chosen allows for a study of the dynamical system without the necessity of selecting explicitly a form for the scalar potential $V\left(\phi\right)$ nor the coupling function $h\left(\phi\right)$ \textit{a priori}, with the explicit forms of these functions being recovered posteriorly.

We have analyzed the structure of the cosmological phase space and extracted the set of fixed points. For any arbitrary values of the scalar field and coupling function, we observe that the phase space of EsGB gravity features the same four types of fixed points as also present in GR, i.e., radiation-dominated, matter-dominated, curvature-dominated, and late-time accelerated cosmological solutions. Furthermore, under an appropriate choice of the coupling function satisfying $\alpha \dot h H < -0.5$, the stability of these fixed points follows the same qualitative behavior as in GR, i.e., radiation dominated solutions are unstable, matter and curvature dominated solutions are saddle points, and late-time cosmic accelerated solutions are stable, the latter corresponding to global attractors in the phase space. Such a structure allows for cosmological solutions starting form a radiation-dominated phase, evolving into a matter-dominated phase, and eventually transitioning into a late-time cosmic accelerated phase, in accordance with the $\Lambda$CDM model. The late-time cosmic acceleration is propelled by the potential $V$, which effectively plays the role of a dark energy component. 

To obtain solutions with cosmological behaviour qualitatively similar to those of the $\Lambda$CDM model, we have followed a reconstruction method in which the GR limit solutions for the matter density parameters and deceleration parameters are introduced into the dynamical system, and then we performed a numerical integration of the dynamical system subjected to appropriate initial conditions. More precisely, we have used the measurements of the Planck satellite to fix the present values of the density parameters, and we have used the weak-field solar system dynamics to fix the present value of the scalar field $\phi$ and its first time derivative. The solutions obtained for the scalar field and the coupling function are finite and regular throughout the entire time evolution independently of the values chosen for the coupling constant $\alpha$, thus implying that a wide variety of EsGB gravity models are capable of reproducing the cosmological behavior of the $\Lambda$CDM model. Such a result, despite indicating a physical relevance of the model, renders the EsGB theory indistinguishable from GR from a cosmological point of view alone.

We note that, in the absence of the scalar potential $V$, late-time cosmically accelerated solutions are not attainable, i.e., the scalar field and its coupling to the Gauss-Bonnet invariant alone can not play the role of a dark energy component and produce a late-time cosmic acceleration, see Appendix \ref{sec:appendix}. However, given the current observational constraints on the value of the cosmological constant in the $\Lambda$CDM model, which induce constraints of the same order of magnitude in the present value of $V$, we argue that this result is not problematic from an astrophysical point of view, where most of the works in EsGB gravity neglect the presence of a potential for the scalar field. The question remains on how the EsGB gravity can be distinguished from GR, given that they seem to be indistinguishable from the point of view of background cosmological evolution only. We will pursue this topic in upcoming works.

\begin{acknowledgments}
The authors would like to acknowledge the anonymous reviewer, who helped them improve the manuscript by pointing out pertinent issues. MASP acknowledges support from the FCT research grants UIDB/04434/2020 and UIDP/04434/2020, and through the FCT project with reference PTDC/FIS-AST/0054/2021 (``BEYond LAmbda''). MASP also acknowledges support from the FCT through the Fellowship UI/BD/154479/2022.
J. L. Rosa is supported by the Project PID2022-138607NBI00, funded by MICIU/AEI/10.13039/501100011033 (``ERDF A way of making Europe" and ``PGC Generaci\'on de Conocimiento"),
\end{acknowledgments}

\appendix

\section{Cosmology in the absence of a scalar potential}\label{sec:appendix}

In other works analyzing the EsGB gravity, it is rather frequent to consider an action of the form of Eq. \eqref{eq:action} but in the absence of the scalar potential $V\left(\phi\right)$. These include works e.g. in the topic of spontaneous scalarization of compact objects \cite{Silva:2017uqg,Cunha:2019dwb}. Thus, we would like to briefly clarify what happens in a cosmological context when the potential $V$ is removed from the analysis. By following the procedure outlined in Secs. \ref{sec:theory} and \ref{sec:dynsys}, the analysis of the critical point in the phase space reveals again a set of five fixed points (see Table \ref{tab:fixedU0} for the coordinates of the fixed points in the phase space, and Table \ref{tab:Stabilityu0} for the corresponding stability analysis). While the radiation-dominated, matter-dominated, and curvature-dominated fixed points, $\mathcal{A}$, $\mathcal{B}$ and $\mathcal{D}$ respectively, are still present in the phase space, the fixed point $\mathcal{C}$, which was previously shown to be an attractor in the phase space and that could lead to a late-time cosmic acceleration, is now dominated by the variable $\Psi$ and corresponds to a linearly growing cosmological solution. This modified phase space structure hints at the fact that late-time cosmologically accelerated solutions can not be obtained in this particular case. 

\begin{figure*}
    \centering
    \includegraphics[scale=0.6]{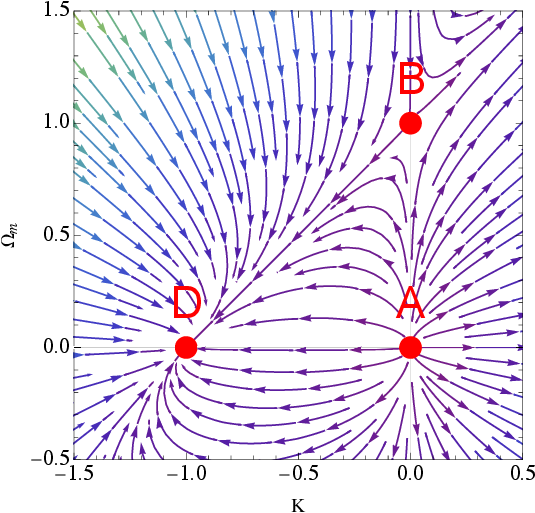}\qquad
    \includegraphics[scale=0.6]{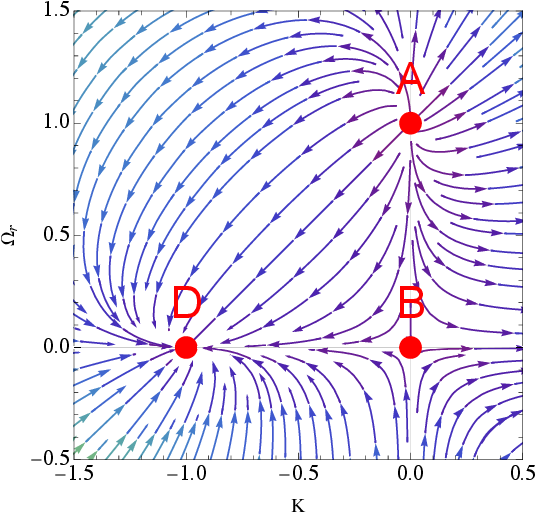}\qquad
    \includegraphics[scale=0.6]{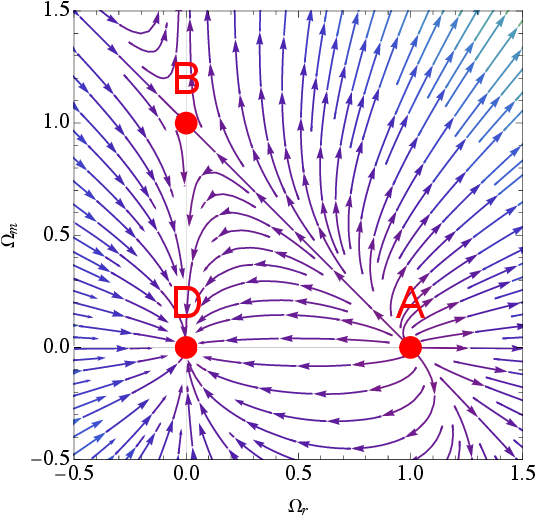}
    \caption{Streamplots of the cosmological phase space for the dynamical system given in Eqs. \eqref{eq:con1} to \eqref{eq:dynTheta}, with $U=0$. The stability analysis of the fixed points represented is given in Table \ref{tab:StabilityGR3}.}
    \label{fig:StreamPlots3}
\end{figure*}

Moreover, we have manipulated the system of Eqs. \eqref{eq:con1}-\eqref{eq:dynTheta} with $U=0$ in the GR limit in the following manner. First, we have used Eq. \eqref{eq:con2} to remove the quantity $Q$ from the system. In addition, Eq. \eqref{eq:con1} is a constraint equation that relates the variables $K$, $\Omega_r$, and $\Omega_m$. This constraint can then be used to remove one of these three quantities from the dynamical system, thus resulting in a simpler two-dimensional system. We have, in turn, removed each of the variables $K$, $\Omega_r$, and $\Omega_m$ from the dynamical system using this constraint and plotted the phase space trajectories for each of the three resulting two-dimensional systems. The streamplots obtained for each of these manipulations are represented in Fig. \ref{fig:StreamPlots3}. In addition, the eigenvalues and stability associated with the three fixed points $\mathcal{A}$, $\mathcal{B}$, and $\mathcal{D}$ is presented in Table \ref{tab:StabilityGR3}. In the absence of the scalar potential, one verifies that points $\mathcal{A}$ and $\mathcal B$ preserve their repeller and saddle behaviors, respectively, whereas point $\mathcal D$ behaves now exclusively as an attractor.

\begin{table*}
\begin{tabular}{c|c c c c c c c} 
 & $K$ & $\Omega_r$ & $\Omega_m$ & $\Theta$ & $\Psi$ & $\Gamma$ & $Q$ \\
\hline $\mathcal{A}$ & 0 & 1 & 0  & 0 & 0 & 0 & 1 \\
$\mathcal{B}$ & 0 & 0 & 1 & 0 & 0 & 0 & $\frac{1}{2}$ \\
$\mathcal{C}$ & 0 & 0 & 0 & 0 & -$\frac{1}{4\alpha}$ & 0 & 0 \\
$\mathcal{D}$ & -1 & 0 & 0 & 0 & ind. & 0 & 0 \\
$\mathcal{E}$ & ind. & 0 & 0 & 0 & $- \frac{1}{4\alpha}$ & 0 & 0
\end{tabular}
\caption{Fixed points arising from the dynamical system of Eqs. \eqref{eq:con1} to \eqref{eq:dynTheta}. The tag ”ind.”
indicates an independent quantity.}
\label{tab:fixedU0}
\end{table*}

\begin{table*}
    \centering
    \begin{tabular}{c| c | c | c| c| c| c| c | c}
         & $\lambda_1$ & $\lambda_2$ & $\lambda_3$ & $\lambda_4$ & $\lambda_5$ & $\lambda_6 $  & Stability\\ 
        \hline $\mathcal{A}$ & 0 & 0 & 2 & 2 & 1 & 0 & Repeller\\ 
        $\mathcal{B}$ & 0 & -1 & 1 & 3/2 & 0 & -1/2  & Saddle \\ 
        $\mathcal{C}$ & 0 & -2 & 0 & 1 & -1 & -1 & Saddle \\ 
        $\mathcal{D}$ & 0 & -2 & 0 & 1 & -1 & -1  & Saddle \\ 
        $\mathcal{E}$ & 0 & -2 & 0 & 1 & -1 & -1 & Saddle \\ 
    \end{tabular}
    \caption{Eigenvalues $\lambda_i$ and stability of the fixed points of the dynamical system in Eqs. \eqref{eq:con1} to \eqref{eq:dynTheta}, with $U=0$.} 
    \label{tab:Stabilityu0}
\end{table*}

\begin{table}
    \centering
    \begin{tabular}{c|c|c}
        $\mathcal A$ & $\mathcal B$ & $\mathcal D$ \\ \hline
       $\begin{matrix}\lambda_1=2\\ \lambda_2=1\end{matrix}$ (R) & $\begin{matrix}\lambda_1=-1\\ \lambda_2=1\end{matrix}$ (S) & $\begin{matrix}\lambda_1=-2\\ \lambda_2=-1\end{matrix}$ (A)  \\ \hline
    \end{tabular}
    \caption{Eigenvalues $\lambda_i$ and stability character of the fixed points of the dynamical system in Eqs. \eqref{eq:con1} to \eqref{eq:dynTheta} with $U=0$, in the GR limit. (A) stands for attractor, (R) stands for repeller, and (S) stands for saddle.}
    \label{tab:StabilityGR3}
\end{table}

\begin{figure*}[htbp]
	\centering
\includegraphics[scale=0.95]{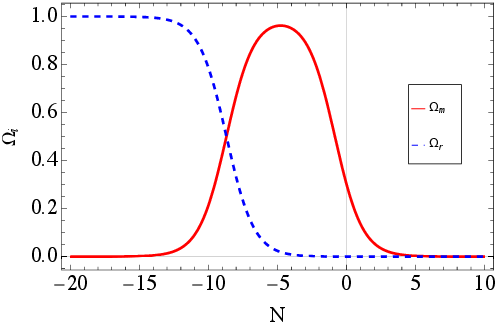}
\includegraphics[scale=0.90]{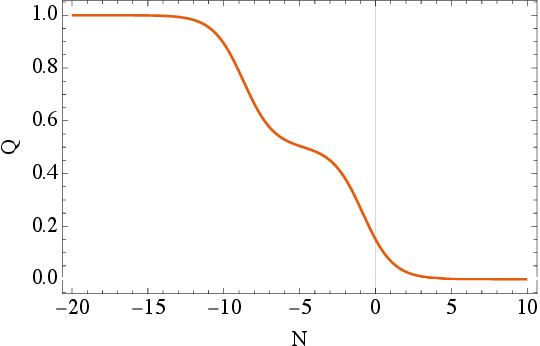}
	\caption{Density parameters $\Omega_r$ and $\Omega_m$ (left panel) and deceleration parameter $Q$ (right panel) as a function of the number of e-folds $N$, obtained through a numerical integration of the dynamical system defined by Eqs. \eqref{eq:con1} to \eqref{eq:dynTheta} in the GR limit $\Phi=\Phi_0=$ constant, $\Theta=0$, and $\Psi = 0$, with $U=0$.}
 \label{fig:parameters0}
\end{figure*}

To confirm that late-time cosmic acceleration can not be achieved in the absence of a potential $V$, we proceed in an analogous way as in Sec. \ref{sec:dynsys}, but setting $U=0$ and removing the dynamical equation for $U$, i.e., Eq. \eqref{eq:dynU}, from the dynamical system. Performing a numerical integration of the resultant dynamical system in this particular case and under the same initial conditions for $\Omega_m$ and $\Omega_r$, see Fig. \ref{fig:parameters0}, one observes that, at late times, the universe asymptotically approaches a linearly expanding $Q=0$ behavior, with both density parameters $\Omega_r$ and $\Omega_m$ approaching zero. Thus, in the absence of a potential $U$ to play the role of the dark energy component in $\Lambda$CDM, the scalar field and the Gauss-Bonnet invariant alone can not propel a late-time cosmic acceleration.

We note that, although such a result could be seen as potentially problematic for the consistency of the theory in several alternative contexts e.g. astrophysics and cosmology, current cosmological observations constrain the value of the cosmological constant to $\Lambda\simeq 10^{-122}M_{\rm pl}^{-2}$, where $M_{\rm pl}$ is the Planck's mass. Thus, in the case of EsGB gravity, the present value of $V\left(\phi\right)$ is constrained to the same value, which is negligible in astrophysical contexts. We thus argue that the necessity of having a potential $V$ to attain compatibility with cosmological observations does not incur in any incompatibilities with the predictions of the theory in the field of astrophysics.



\end{document}